\documentclass[10pt,fleqn]{article}

\usepackage{epsfig}
\usepackage{subfigure}
\newcommand{\beq}{\begin{equation}}
\newcommand{\eeq}{\end{equation}}
\def\lsim{\mathrel{\rlap{\lower4pt\hbox{\hskip1pt$\sim$}}
    \raise1pt\hbox{$<$}}}         
\def\gsim{\mathrel{\rlap{\lower4pt\hbox{\hskip1pt$\sim$}}
    \raise1pt\hbox{$>$}}}         
\def\esim{\mathrel{\rlap{\raise2pt\hbox{$\sim$}}
    \lower1pt\hbox{$-$}}}         

\begin{document}

\begin{titlepage}
\pagestyle{empty}
\title{Full One-Loop Calculation of \\ Neutralino Annihilation\\ into Two Photons}
\author{Lars Bergstr{\"o}m\thanks{lbe@physto.se} \\
 Department of Physics, Stockholm University,\\
 Box 6730, SE-113~85~Stockholm, Sweden
\and
 Piero Ullio\thanks{piero@teorfys.uu.se} \\
 Department of Theoretical Physics, Uppsala University,\\
 Box 803, SE-751 08 Uppsala, Sweden}
\date{June 12, 1997}

\maketitle
\smallskip

\begin{abstract}
For the first time, a full one-loop calculation of the 
process $\chi\chi\to 2\gamma$ is
performed, where $\chi$ is the lightest neutralino in the minimal 
supersymmetric extension of the Standard Model. This process is of 
interest for dark matter detection, since it would give a sharp
$\gamma$ ray line with $E_\gamma=m_\chi$. We improve upon and correct 
published formulas, and give cross sections for supersymmetric models with
$\chi$ masses between 30 GeV and several TeV. We find a new contribution, 
previously neglected,
which enhances the $2\gamma$ rate for TeV higgsinos by up to an order 
of magnitude. As a byproduct, we obtain a new expression for the related
process $\chi\chi\to 2$ $gluons$, which on the other hand is 
generally smaller 
than previously calculated.

There has been a recent claim that evidence for a 3.5 TeV 
higgsino annihilating into a $\gamma$ line may already exist from 
balloon emulsion and Air Cherenkov Telescope data. We 
comment on attractive features and problems with this interpretation.
\end{abstract}
\end{titlepage}

\section{Introduction}\label{sec:intro}

Most extensions of the Standard Model of particle physics contain 
 supersymmetry in one or another form. In particular, if 
gravity is unified with the other fundamental forces, it seems that 
supersymmetry is a necessary ingredient. The known particles of 
nature seem to be unstable, with very few exceptions. Of course, the 
stable ones have a specific significance since they have survived 
since the times immediately following the Big Bang, and today form 
the visible structures (clusters, galaxies, stars, planets\ldots) in 
the Universe. 

If there exist 
stable particles in the supersymmetric sector, those should also have 
survived, and could give a contribution to the total mass density of 
the Universe in a way which is calculable once the supersymmetric 
model is specified. Models which could contain stable particles 
are usually endowed with a symmetry, $R$-parity, which guarantees that 
the lightest supersymmetric particle is stable (for a review, see
\cite{jkg}). If this particle is electrically neutral, as is the case 
for the neutralino (a linear combination of the superpartners of the 
photon, the $Z$ boson and two neutral $CP$-even Higgs bosons), 
its weak interaction strength provides a relic 
density which can be within an order of magnitude of the critical 
density of the Universe,
\beq
\rho_{crit}=1.9\cdot 10^{-29}h^2\ {\rm g}/{\rm cm}^{3},
\eeq
where $h$ is the Hubble parameter in units of $100$ 
km\,s$^{-1}$Mpc$^{-1}$. Present measurements \cite{freedman, 
kochanek} indicate that $h$ is in the range $0.5-0.7$.
It is customary when discussing energy density to normalize to the 
critical density through the dimensionless quantity $\Omega$:
\beq
\Omega={\rho\over \rho_{crit}}.
\eeq
With the present value of $h$, the critical density $\Omega=1$ 
corresponds to 
$\Omega h^2=0.25-0.5$. 

From dynamical estimates of the total mass density on cluster and 
supercluster scales, it is found that $\Omega_{tot}>0.4$ \cite{dekel}.
An analysis based on supernovas of type Ia as standard candles 
\cite{saul} is not inconsistent with $\Omega=1$, a value which has a 
strong theoretical preference since it is predicted in generic models 
of inflation. The contribution $\Omega_{b}$ from ordinary 
matter (mainly baryons) is limited by nucleosynthesis arguments to be 
less than around $0.1$ \cite{copi}. Also, structure formation from the 
primordial fluctuations 
seen in the microwave background seems to be 
problematic unless there is a considerable amount of dark matter 
\cite{kolbturner}.

The ``missing mass'' problem found in the analysis of galactic 
rotation curves \cite{zwicky} may thus have as its solution the 
existence of a relic population of neutral massive particles residing 
in the galactic halos. We will assume that these particles are 
neutralinos $\chi\equiv\tilde\chi^0_{1}$, the lightest supersymmetric 
particle in the minimal 
supersymmetric extension of the Standard Model.

\section{Detection methods}

Several methods have been discussed  to detect neutralinos (for a 
comprehensive list of reference for all methods mentioned here, see 
\cite{jkg}). One 
possibility is to register scattering of neutralinos directly in 
sensitive terrestrial detectors, well shielded from radioactive and 
other backgrounds. After several years of development, some of these 
experiments now have a sensitivity large enough to rule out some 
supersymmetric models which otherwise satisfy all accelerator and 
other constraints \cite{bg,bott}.

Other, more indirect, methods utilize the fact that neutralinos in 
the halo (or those accumulated gravitationally in massive bodies like 
the Earth and the Sun) may annihilate, resulting in ordinary particles 
which may be detected. For example, antiprotons and positrons which 
are not produced in large quantities by cosmic rays, would be 
produced in annihilations in the halo and would be continuously 
injected and 
``stored'' during around $10^7$ years which is the containment time 
of cosmic rays in the galaxy. Although the fluxes could be of 
measurable size the problem is that they are generally not much 
larger than the cosmic-ray induced fluxes, which necessitates a 
difficult model-dependent background subtraction. Also, since the 
spatial and energy distributions of these antiparticles are rather 
featureless, the lack of definitive signature is a problem. This is 
also true for annihilation into particles which give secondary gamma 
rays. Although this could give a broad feature in the spectrum up to 
the neutralino mass, it is in most supersymmetric models
 too weak to be convincing \cite{hansuno}.

A much better signature is given by neutralino annihilation into neutrinos near 
the centre of the Earth or the Sun. Since the ordinary solar neutrinos 
only have at most MeV energies, a multi-GeV neutrino signal from the 
Sun (or the Earth) would be an unmistakeable signal. If the neutrino 
detector has an angular resolution better than a few degrees, the mass 
of the annihilating particles can be determined with reasonable 
accuracy \cite{eg}. Also this method has already given useful 
constraints on supersymmetric models of dark matter \cite{beg}, and 
has a large future potential as large new neutrino telescopes are 
being built \cite{gaisser}. 

Finally, an excellent signature is provided by the annihilation of 
neutralinos in the 
halo into 
two-body final states containing a photon, such as $\gamma\gamma$. This 
process was first proposed by Bergstr\"om and Snellman \cite{lb1}, who 
also first estimated the related two-gluon process (for photinos).
Other early references include \cite{rudaz,bouquet,giudice}.
Since the annihilating neutralinos move with galactic 
velocities, $v/c\sim 10^{-3}$, the result is a gamma ray line of 
phenomenal peak-to-width ratio. This signature has no conceivable 
background from known astrophysical sources. To optimally utilize 
this sharp signal peak, detectors with energy resolution at the 
percent level and below should ideally be used. In the region 30 - 100 GeV, 
where ``light'' neutralinos (mostly rather pure binos) are likely to reside,
a detector has to be placed in orbit (above the atmosphere) not to be 
flooded with cosmic-ray produced atmospheric gammas. This led to 
a (temporary?) impasse for this idea, since photon detectors with high 
energy resolution are generally very heavy and therefore costly 
to launch. 

It was subsequently realised, in work by Urban et al \cite{urban}, that a 
certain type of ground-based detector (Air Cherenkov Telescopes, ACT) 
could be used for searching for gamma lines from very heavy 
neutralinos (several hundred GeV to the TeV region). Here the limited 
energy resolution (10 \%, at best) is partly offset by the very large 
collection areas possible ($10^{4}$ to $10^5$ m$^2$). The rates for 
TeV neutralinos (generally higgsinos, since binos of such masses would 
overclose the Universe) were calculated in \cite{lbe1}. In this 
paper, the process $\chi\chi\to Z\gamma$ was also proposed as an important
channel, and a
lower bound based on unitarity for high-mass higgsino annihilation into 
$\gamma\gamma$ and $Z\gamma$ was computed. 
In \cite{juka}, a more complete calculation 
of the $\gamma\gamma$ process was presented, and the phenomenology 
was also worked out in \cite{bottino} (although in none of these TeV 
neutralinos were considered).

Since there is a new generation of Air Cherenkov telescopes being 
planned \cite{ong}, and also a new ambitious satellite project 
\cite{glast}, we have considered it timely to revisit the 
$\gamma\gamma$ process, performing for the first time a full one-loop 
calculation improving on the estimates of previous work. As a 
byproduct, we also recompute the two-gluon process, correcting a 
numerically significant error in \cite{drees}. We give analytical 
expressions for all the amplitudes and give results of scans in 
supersymmetric parameter space for several sets of models.

During the final stages of this work, we have come upon a recent 
paper \cite{strausz}, where it is claimed that evidence for a TeV 
higgsino annihilating into a $\gamma$ line may already exist from 
balloon emulsion and ACT data. We will 
comment on attractive features and problems with this interpretation.

It is important to note that all the 
different methods to detect supersymmetric dark matter are 
complementary in many respects. Each process has certain regions in 
parameter space where it dominates. In fact, for TeV higgsinos, the 
$\gamma$ line process may probe supersymmetric models out of reach
for accelerators in the forseeable future (for other possible 
methods to detect heavy higgsinos, see
\cite{beg2}).

\section{Cross section}
\label{sec:CRSEC}

We compute the full expression for the annihilation cross section of 
the 
process
\beq
\tilde{\chi}^{0}_{1} + \tilde{\chi}^{0}_{1} \rightarrow \gamma 
+\gamma 
\eeq
in the limit of vanishing relative velocity of the neutralino pair. 
This 
result can be applied to neutralinos annihilating in the galactic 
halo, where
it is estimated that  massive dark matter particles move with a 
velocity 
of the order $v/c \sim$ 10$^{-3}$. The outgoing photons will then be 
nearly 
monochromatic, with an energy equal to the mass of ${\chi}^{0}_{1}$. 
Furthermore, for very low relative velocity, the neutralino pair must 
be in 
a S wave state with pseudoscalar quantum numbers; we can therefore 
project 
out of the amplitude the $^{1}$S$_{0}$ state, using the 
projector~\cite{kuhn}
\beq
\mathcal{O}_{Ps}=- \frac{M_{\chi}}{\sqrt{2}} \; \gamma^{5}
 \left(1-\frac{\,\not\!p}{M_{\chi}} \right)
\eeq
where $M_{\chi}$ is the mass of \(\tilde{\chi}^{0}_{1}\) and 
$p = ( M_{\chi},\vec{0}\,)$ its momentum. Analyzing the Dirac 
structure of 
the Feynman diagrams we can draw for this process, we can exclude a 
priori 
those diagrams which cannot generate a pseudoscalar state.

\begin{figure}
 \centering
 \mbox{\subfigure{\epsfig{file=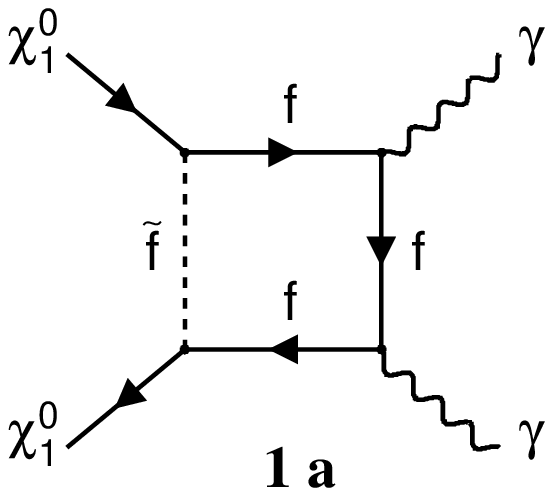,width=3cm}}\quad
       \subfigure{\epsfig{file=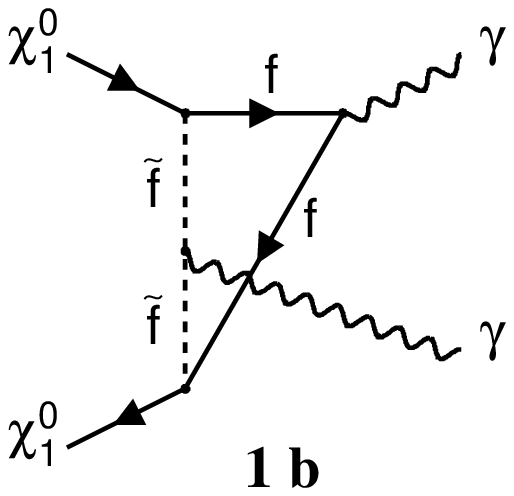,width=3cm}}\quad
       \subfigure{\epsfig{file=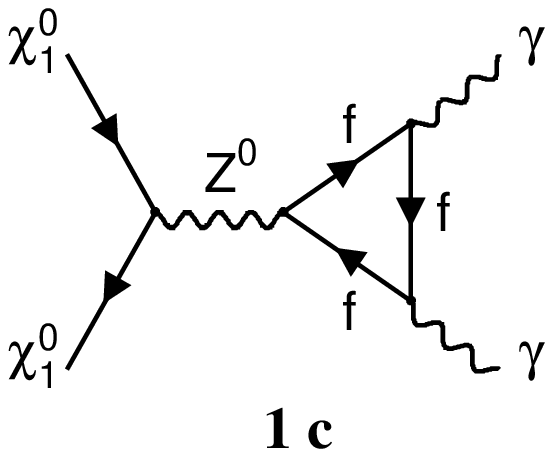,width=3cm}}\quad
       \subfigure{\epsfig{file=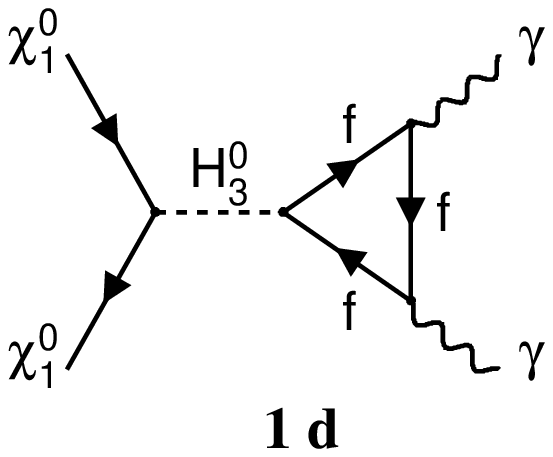,width=3cm}}}
 \mbox{\subfigure{\epsfig{file=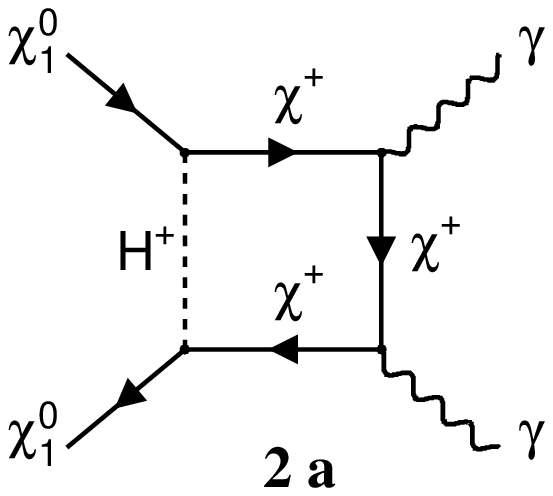,width=3cm}}\quad
       \subfigure{\epsfig{file=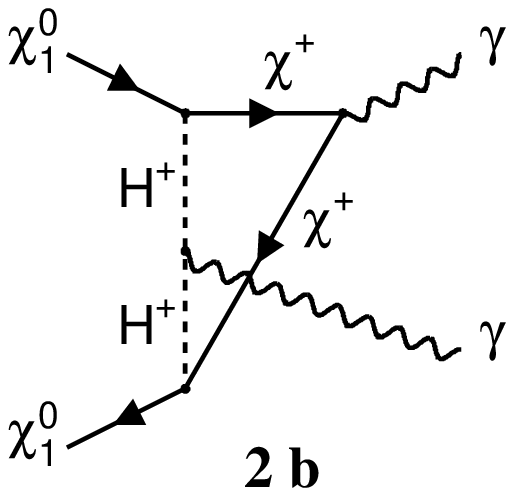,width=3cm}}\quad
       \subfigure{\epsfig{file=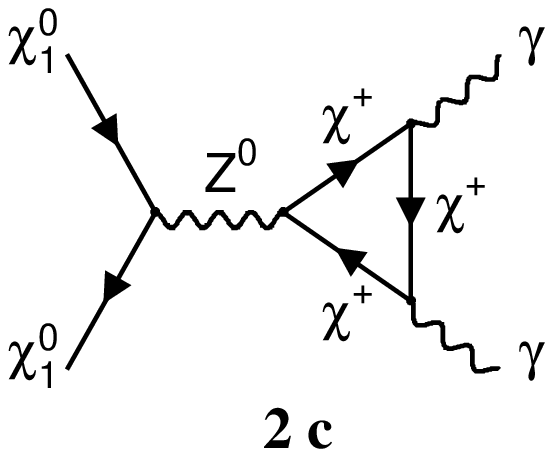,width=3cm}}\quad
       \subfigure{\epsfig{file=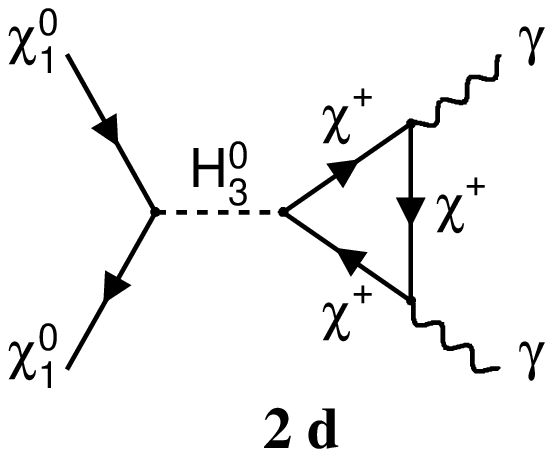,width=3cm}}}
 \mbox{\subfigure{\epsfig{file=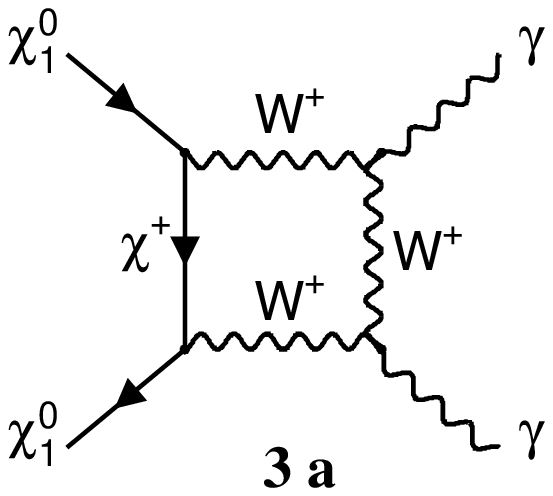,width=3cm}}\quad
       \subfigure{\epsfig{file=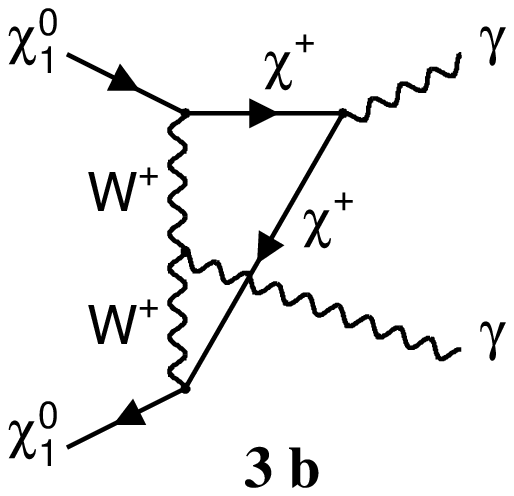,width=3cm}}\quad
      \subfigure{\epsfig{file=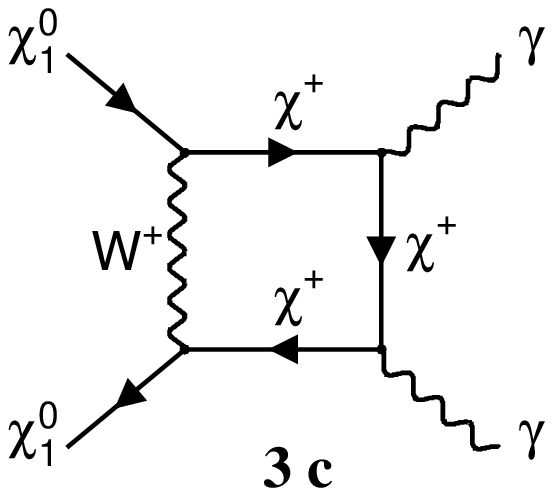,width=3cm}}}
 \mbox{\subfigure{\epsfig{file=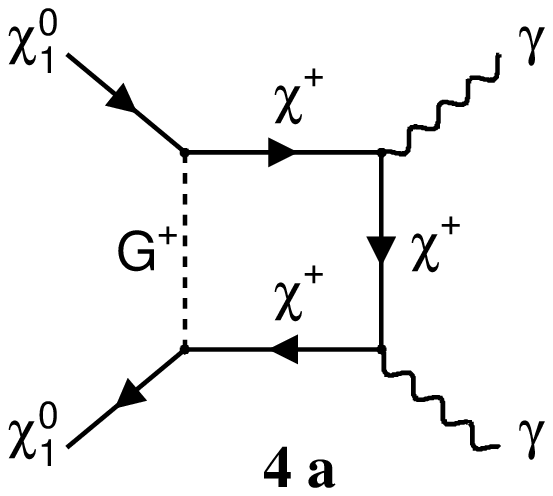,width=3cm}}\quad
       \subfigure{\epsfig{file=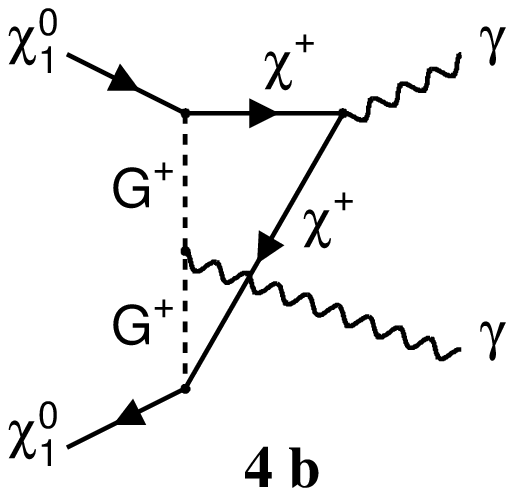,width=3cm}}}
\caption{Feynman diagrams contributing, at one loop level, to neutralino 
annihilation into two photons. Diagrams with exchanged initial and final 
states are not shown.}\label{fig:fig1}
\end{figure}

As suggested in Ref.~\cite{lbe1}, it is convenient to compute the W 
boson and 
unphysical Higgs loop diagrams choosing the non linear gauge defined in 
Ref.~\cite{fujikawa}, which is a slight variant of the usual linear 
R-gauge (or 't Hooft gauge). In this particular gauge the 
$W^{\pm}\,G^{\mp}\,\gamma$ vertex vanishes and some of the box 
diagrams which,
in the linear R-gauge, may give a contribution to the amplitude are 
not 
present. In addition the $W^{\pm}\,W^{\mp}\,\gamma$ vertex and the W 
propagator assume simpler forms. We find that the Feynman diagrams we 
have to include in the computation, at one loop level, are those 
shown in 
Fig.\,\ref{fig:fig1}. 

There is another property that introduces a further main 
simplification. In any process of annihilation of particles at rest, 
the four 
point loop integrals one obtain from each box diagram applying the 
Feynman rules, can always be rewritten as a linear combination of 
three-point integrals. This is possible because, for each integral, one can 
find a 
linear combination of the four factors in the denominator which is 
independent
of the momentum flowing in the loop. It follows that we do not have 
any four-point integral in the computation of the amplitude.

The amplitude of the annihilation process can be factorized in the 
form
\beq
\mathcal{A} =\frac{e^2}{2 \sqrt{2} \; \pi^2}
 \epsilon\left(\epsilon_{1},\epsilon_{2},k_{1},k_{2} \right)
 \;\tilde{\mathcal{A}}
\eeq
where $\epsilon_{1}$, $\epsilon_{2}$ and $k_{1}$, $k_{2}$ are 
respectively 
the polarization tensors and the momenta of the two outgoing photons.
The cross section is then given, as a function of 
$\tilde{\mathcal{A}}$, by the formula
\beq
 v\sigma_{2\gamma} = \frac{\alpha^2 M^2_{\chi}}{16 \pi^3} \left|\; 
\tilde{\mathcal{A}}\; \right|^{2}\;\;\;.  \label{eq:sigmav}
\eeq 

We write the total amplitude as the sum of the contributions obtained 
from the
four different classes of diagrams:
\begin{eqnarray*}
\tilde{\mathcal{A}}=\tilde{\mathcal{A}}_{f\tilde{f}}+
 \tilde{\mathcal{A}}_{H^+}+\tilde{\mathcal{A}}_{W}+\tilde{\mathcal{A}}_{G}.
\end{eqnarray*}
The single contributions are listed below. In each of them we 
separate 
real and imaginary parts. To facilitate a comparison, we follow as 
closely as possible the notation of \cite{juka}.

1) Contribution of the fermion-sfermion loop diagrams
(1a-1d in Fig.\,\ref{fig:fig1}). The sum over $f$ includes the quarks and the 
charged leptons, the sum over $\tilde{f}$ the corresponding sfermion 
in both 
chiral states.
\begin{eqnarray}
Re\,\tilde{\mathcal{A}}_{f\tilde{f}} & = &
\sum_{f\tilde{f}} c_f \frac{e_{f}^2}{M^2_{\chi}} \; \left[ 
\frac{1}{2}\; 
 \frac{b\;S_{f\tilde{f}}+\sqrt{a\,b}\;D_{f\tilde{f}}}{1+a-b} \;I_{1} 
 \left( a,b \right)+ \frac{1}{2}\; \frac{S_{f\tilde{f}}}{1-b} \;I_{2}
 \left( a,b \right) \right. \nonumber\\
  && +\left. 
\left(\frac{b\;S_{f\tilde{f}}+\sqrt{a\,b}\;D_{f\tilde{f}}}
   {1+a-b}-\frac{1}{2}\;\frac{b\;S_{f\tilde{f}}}{1-b}\right)\;I_{3}
   \left( a,b \right) \right] \nonumber\\
  && +\sum_{f} c_f\frac{e_{f}^2}{M^2_{\chi}} \; \left[ \left(  
   \frac{m^2_{f}\;G_{Zf}}{4\,m^2_{Z}} -\frac{m_{f}\,M_{\chi}\;
   G_{H^{0}_{3} f}}{4\,\left(4\,M^2_{\chi}-m^2_{H^{0}_{3}} \right)}
   \right)\;I_{1}\left(a,b\right)\right] \label{reff}
\end{eqnarray}

\begin{eqnarray}
Im\,\tilde{\mathcal{A}}_{f\tilde{f}} & = & -\pi\;
 \sum_{f}  c_f \frac{e_{f}^2}{M^2_{\chi}} \; \left[ \sum_{\tilde{f}} 
 \left(\frac{1}{2}\; 
\frac{b\;S_{f\tilde{f}}+\sqrt{a\,b}\;D_{f\tilde{f}}}
 {1+a-b} \right) \;  +  \frac{m^2_{f}\;G_{Zf}}{4\,m^2_{Z}} \right. 
\nonumber\\
 && \left. -\frac{m_{f}\,M_{\chi}\;G_{H^{0}_{3} 
f}}{4\,\left(4\,M^2_{\chi}
  -m^2_{H^{0}_{3}} \right) }\; \right] \,\cdot \, \log \left( 
  \frac{1+\sqrt{1-b/a}}{1-\sqrt{1-b/a}} \right) \theta 
\left(1-m^2_{f}\,
   /\,M^2_{\chi} \right)\label{eq:imsquark} 
\end{eqnarray}

where $m_{Z}$ and $m_{H^{0}_{3}}$ are the masses of the Z boson and 
of the Higgs pseudoscalar $H^{0}_{3}$, and we have defined:
\begin{eqnarray*}
a=\frac{M^2_{\chi}}{M^2_{\tilde{f}}} && 
b=\frac{m^2_{f}}{M^2_{\tilde{f}}}
\end{eqnarray*}
\begin{eqnarray*}
S_{f\tilde{f}}=\frac{1}{2}\;\left(g^L_{\tilde{f}f1}\; g^{L\,\ast}_
 {\tilde{f}f1}+g^R_{\tilde{f}f1}\; g^{R\,\ast}_{\tilde{f}f1} \right) 
 && D_{f\tilde{f}}=\frac{1}{2}\;\left( g^L_{\tilde{f}f1}\; g^{R\,\ast}
  _{\tilde{f}f1}+g^R_{\tilde{f}f1}\; g^{L\,\ast}_{\tilde{f}f1} \right)
\end{eqnarray*}
\begin{eqnarray*}
G_{Zf}=\frac{g 
\;T^3_{f}}{\cos\;\theta_{W}}\;\left(g^L_{Z11}-g^R_{Z11} \right)
 && G_{H^{0}_{3} f}=\left(g^R_{H^{0}_{3}11}-g^L_{H^{0}_{3}11} \right) 
\,
  \left(g^L_{H^{0}_{3}ff}-g^R_{H^{0}_{3}ff} \right)\;\; .
\end{eqnarray*} 
The index 1 is referred to \(\tilde{\chi}^{0}_{1}\), $e_{f}$ is the 
charge of 
the fermion in unit of the electron charge ($-e$), $c_f$ is the color 
factor 
equal to 3 for quarks and to 1 for leptons, $T^3_{f}$ is the third 
component 
of the weak isospin equal to +1/2 for f=u,c,t and to -1/2 for 
f=e,$\mu$,$\tau$,d,s,b.    

2) Contribution of the chargino-Higgs boson loop diagrams (2a-2d in 
Fig.\,\ref{fig:fig1}). 
The sum over $\chi^+_i$ includes the two chargino eigenstates.
\begin{eqnarray}
Re\,\tilde{\mathcal{A}}_{H^+} & = &
\sum_{\chi^+_i} \frac{1}{M^2_{\chi}} \; \left[ \frac{1}{2}\; \frac{b\;
 S_{\chi H}+\sqrt{a\,b}\;D_{\chi H}}{1+a-b} \;I_{1} \left( a,b 
\right)+ 
 \frac{1}{2}\; \frac{S_{\chi H}}{1-b} \;I_{2}\left( a,b \right)  
\right. 
 \nonumber \\
  && \left. + \left(\frac{b\;S_{\chi H}+\sqrt{a\,b}\;D_{\chi 
H}}{1+a-b}-
   \frac{1}{2}\;\frac{b\;S_{\chi H}}{1-b}\right)\;I_{3}\left( a,b 
\right) 
   \right. \nonumber \\
  && \left. + \left(  \frac{M^2_{\chi^+_i}\;G_{Z\chi}}{4\,m^2_{Z}} 
   -\frac{M_{\chi^+_i}\,M_{\chi}\;G_{H^{0}_{3} 
\chi}}{4\,\left(4\,M^2_{\chi}
   -m^2_{H^{0}_{3}} \right)}\right)\;I_{1}\left(a,b\right)\right] 
\end{eqnarray}

\beq
Im\,\tilde{\mathcal{A}}_{H^+}\;=\;0
\eeq
where we have defined:
\begin{eqnarray*}
a=\frac{M^2_{\chi}}{m^2_{H^+}} & & b=\frac{M^2_{\chi^+_i}}{m^2_{H^+}}
\end{eqnarray*} 
\begin{eqnarray*}
S_{\chi H}=\frac{1}{2}\;\left(g^L_{H^+ 1i}\; g^{L\,\ast}_{H^+ 1i}
 +g^R_{H^+ 1i}\; g^{R\,\ast}_{H^+ 1i} \right) 
 && D_{\chi H}=\frac{1}{2}\;\left( g^L_{H^+ 1i}\; g^{R\,\ast}_{H^+ 1i}
  +g^R_{H^+ 1i}\; g^{L\,\ast}_{H^+ 1i} \right)
\end{eqnarray*}
\begin{eqnarray*}
G_{Z \chi}=\left(g^R_{Zii}-g^L_{Zii} \right) 
\,\left(g^L_{Z11}-g^R_{Z11} 
 \right) && G_{H^{0}_{3} 
\chi}=\left(g^R_{H^{0}_{3}11}-g^L_{H^{0}_{3}11} 
 \right) \,\left(g^L_{H^{0}_{3}ii}-g^R_{H^{0}_{3}ii} \right)
\end{eqnarray*} 
and the index i is referred to \(\tilde{\chi}^{+}_{i}\) .

3) Contribution of the chargino-W boson loop diagrams (3a-3c in 
Fig.\,\ref{fig:fig1}).
\begin{eqnarray}
Re\,\tilde{\mathcal{A}}_{W} & = &
\sum_{\chi^+_i} \frac{1}{M^2_{\chi}} \; \left[ 2\; 
\frac{\left(a-b\right)\;
 S_{\chi W}}{1+a-b} \;I_{1} \left( a,b \right)+\frac{S_{\chi W}-2 
\sqrt{a}\;
 D_{\chi W}}{1-a-b} \;I_{1} \left( a,1 \right) \right. \nonumber \\
  &&\left. + \left( 2\;\frac{S_{\chi W}-2 \sqrt{a}\;D_{\chi 
W}}{1-a-b}-
   \frac{3\,S_{\chi W}-4 \sqrt{a}\;D_{\chi W}}{1-b}  \right)\;I_{2}
   \left( a,b \right) \right. \nonumber \\
  && \left.+ \left( \frac{\left(2+b\right)\,S_{\chi W}-4 \sqrt{a}\;
   D_{\chi W}}{1-b} -2\;\frac{\left( 1-a+b \right)\; S_{\chi 
W}}{1+a-b} 
   \right)\;I_{3}\left( a,b \right) \right] 
\end{eqnarray}

\begin{eqnarray}
Im\,\tilde{\mathcal{A}}_{W} & = &
 -\pi\;\sum_{\chi^+_i} \frac{1}{M^2_{\chi}} \; \left( 2\; 
  \frac{\left(a-b\right)\;S_{\chi W}}{1+a-b} \right) \cdot \nonumber 
\\
 &&\cdot \, \log \left( \frac{1+\sqrt{1-b/a}}{1-\sqrt{1-b/a}} \right) 
  \theta \left(1-m^2_{W}\,/\,M^2_{\chi} \right) \label{imw}
\end{eqnarray}

where in this case:
\begin{eqnarray*}
a=\frac{M^2_{\chi^0_1}}{M^2_{\chi^+_i}} && 
b=\frac{m^2_{W}}{M^2_{\chi^+_i}}
\end{eqnarray*}
\begin{eqnarray*}
S_{\chi W}=\frac{1}{2}\;\left(g^L_{W1i}\; 
g^{L\,\ast}_{W1i}+g^R_{W1i}\; 
 g^{R\,\ast}_{W1i} \right) && D_{\chi W}=\frac{1}{2}\;\left( 
g^L_{W1i}\; 
 g^{R\,\ast}_{W1i}+g^R_{W1i}\; g^{L\,\ast}_{W1i} \right)\;\; .
\end{eqnarray*}

4) Contribution of the chargino-unphysical Higgs loop diagrams (4a-4b in 
Fig.\,\ref{fig:fig1}).
\begin{eqnarray}
Re\,\tilde{\mathcal{A}}_{G} &=&
\sum_{\chi^+_i} \frac{1}{M^2_{\chi}} \; \left[ -\frac{1}{2} \; 
 \frac{S_{G\chi}+ \sqrt{a}\;D_{G\chi}}{1-a-b} \;I_{1} \left( a,1 
\right) 
 \right. \nonumber \\
  && \left. + \left(\frac{1}{2}\;\frac{S_{G\chi}}{1-b} 
-\frac{S_{G\chi}+ 
   \sqrt{a}\;D_{G\chi}}{1-a-b}  \right)\;I_{2}\left( a,b \right) 
-\frac{1}{2}
   \;\frac{b\; S_{G\chi}}{1-b}\;I_{3}\left( a,b \right) \right] 
\end{eqnarray}

\beq
Im\,\tilde{\mathcal{A}}_{G}\;=\; 0
\eeq

where a and b are the same as those defined for the chargino-W boson 
contribution and
\begin{eqnarray*}
S_{G\chi}=\frac{1}{2}\;\left(g^L_{G1i}\; 
g^{L\,\ast}_{G1i}+g^R_{G1i}\; g^{R\,\ast}_{G1i} \right) && 
D_{G\chi}=\frac{1}{2}\;\left( g^L_{G1i}\; 
g^{R\,\ast}_{G1i}+g^R_{G1i}\; g^{L\,\ast}_{G1i} \right)\;\; .
\end{eqnarray*}

The coupling constants for left and right chiral states, $g^{L}$ and 
$g^{R}$, are written in the notation adopted in the PhD Thesis of 
Edsj\"{o}~\cite{joakim}; all of them are defined therein, with the 
exception of the unphysical Higgs couplings, which we rewrite from 
Ref.~\cite{haber-wyler} in the conventions of \cite{joakim}:
\begin{eqnarray}
  g^L_{G1i} & = & - g \sin \beta \left[
    Z_{14}^* V_{i1}^* + \sqrt{\frac{1}{2}} 
    (Z_{12}^* + Z_{11}^* \tan \theta_W) V_{i2}^*
    \right] \\
  g^R_{G1i} & = & + g \cos \beta \left[
    Z_{13} U_{i1} - \sqrt{\frac{1}{2}} 
    (Z_{12} + Z_{11} \tan \theta_W) U_{i2}
    \right] \varepsilon_i
\end{eqnarray}
The functions $I_{1}\left( a,b \right)$, 
$I_{2}\left( a,b \right)$ and $I_{3}\left( a,b \right)$, which arise 
from 
the loop integrations, are specified in Appendix A. $I_{1}\left( a,b 
\right)$
is the well known three point function that appears in triangle 
diagrams; it 
is an analytic function of a/b. $I_{2}\left( a,b \right)$ and 
$I_{3}\left( a,b \right)$ are given both in form of integrals over a 
Feynman parameter x, and may be expressed 
in terms of dilogarithms. It is numerically 
very 
easy to compute them even in the integral form as, for any physically 
interesting value of the parameters a and b, the integrands are 
smooth 
functions of x. In the real parts there are apparently poles 
for $a - b = 1$, $b = 1$, $a + b = 1$: we have verified that, in each 
formula,
a cancellation among the different contributions takes place for each 
of 
these spurious poles and therefore we do not have to deal with any 
numerical 
problem.

The results for the imaginary parts have been checked applying the 
Cutkosky cutting 
rules to the corresponding diagrams. Eq.\,~(\ref{reff}) reproduces 
equations
(2.5) in Ref.~\cite{drees} and (2.5) in Ref.~\cite{juka} except for
one small, but numerically significant, error in both of 
these.\footnote{In the formula (2.5) of \cite{drees}, we think that
the argument 
of the denominator of the second logarithm should be $b-a(1-x^2)$ 
instead of $b+a(1-x^2)$. This is plausible on general grounds, since
it is precisely the zero of this expression in the integrand which gives
an imaginary part when $m_\chi >m_f$. It seems this error has propagated
to the computer code of \cite{jkg}.}
 As we will briefly sketch in 
Section \ref{sec:gluons}
 this discrepancy may have led, for some SUSY models, to an 
overestimate of 
the cross section of the annihilation of neutralinos into two gluons. 
In Eq.\,~(\ref{imw}), fixing the overall sign through 
Eq.\,~(\ref{eq:imsquark}), we do not agree with the sign of the 
result quoted in Ref.\,~\cite{juka}.  This sign difference is numerically 
important in cases where the amplitudes are of the same order of 
magnitude.

\section{Results and Discussion}\label{sec:results}

We have calculated the $\chi\chi\to 2\gamma$ rates for a broad selection
of supersymmetric models. All experimental constraints on the
parameters have been checked, and only non-excluded models have been
treated. (For a detailed explanation of the method of computation
and the parametrization of models, see \cite{joakim}.)

The calculation of the relic density $\Omega h^2$ has been done using
the full machinery of exact treatment of resonances 
and thresholds \cite{paolograc}, loop corrections to  masses and
vertices
\cite{drees2} and coannihilations \cite{coann}.
We are only interested in models which have a chance to at least 
explain the galactic halos, so we keep in all the graphs presented 
here only models which satisfy $0.025<\Omega h^2<1$.

The detection rate of galactic gamma rays from neutralino annihilations
depends on two different parts. One concerns the particle physics calculation
of the $2\gamma$ cross section for a given set of supersymmetric model
parameters, and also the calculation of the relic abundance of neutralinos
in standard Big Bang cosmology. The second part concerns the distribution
of neutralinos in the galactic dark matter halo, which is a difficult problem
of galactic structure formation. Unfortunately, all detection predictions are
very sensitive to the poorly known dark matter mass distribution in the 
halo, since the $\gamma$ line
flux from a volume element $dV$ of density $\rho_\chi$ is given by
\beq
d\,\Phi_\gamma= 2\,\frac{v\sigma_{2\gamma}\;\rho_\chi^2}
{4\,\pi\,d^2\;M_\chi^2}\;d\,V,
\eeq
where $d$ is the distance to the volume element and 
the factor of 2 comes because there are two $\gamma$s per annihilation. 
To obtain the full rate, one has to integrate this expression over all
the halo (or, for a particular direction, over the line-of-sight). 

We start by 
discussing the halo-model independent
quantities, which can be used for any particular halo dark matter 
distributions. Later we discuss the expected rates for some examples of
halo models.

\begin{figure}
\begin{center}
 \mbox{\subfigure{\epsfig{file=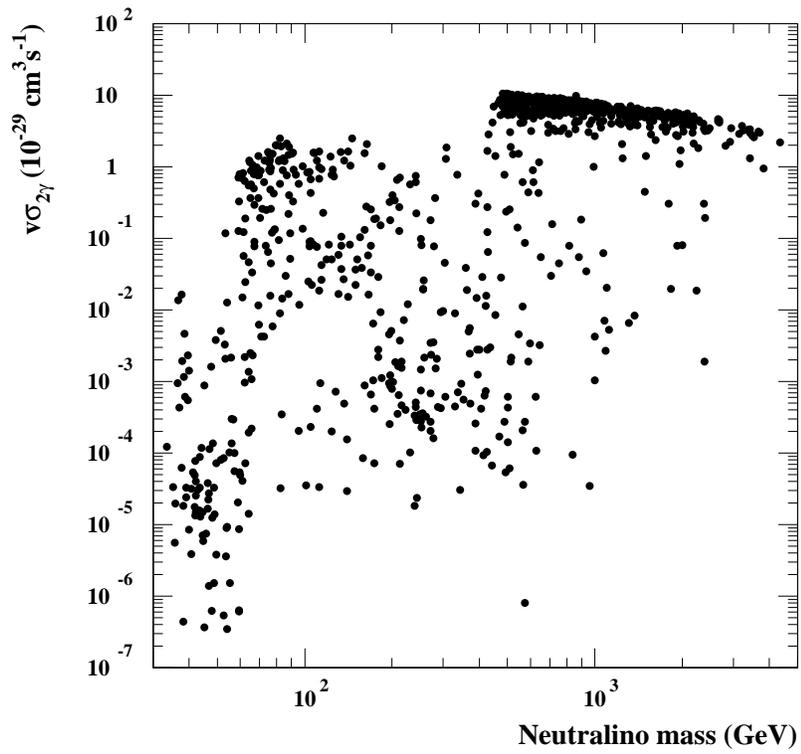,width=12cm}}}
\caption{Annihilation rate of neutralinos into 2$\;\gamma$ for a broad scan 
in supersymmetric parameter space (the parameter range is identical to that 
of the ``generous scan'' of Table 2.2 in~[34,30]).}\label{fig:fig2}
\end{center}
\end{figure}

In Fig.\,\ref{fig:fig2} we show the prediction for the annihilation
rate into two photons, $v\sigma_{2\gamma}\equiv 
v\sigma(\tilde{\chi}^{0}_{1} + \tilde{\chi}^{0}_{1} \rightarrow \gamma 
+\gamma)$ for a broad scan in supersymmetric parameter space (the
parameter range is identical to that of the ``generous scan'' of Table 2.2 in
\cite{coann, joakim}). As can be seen,  for the low-mass region the rate varies
over many orders of magnitude. In the most favourable cases, it is of
the order of a few times $10^{-29}$ cm$^3$s$^{-1}$, whereas it can be as low
as $10^{-6}$ of that value for other models. This wide spread of 
predictions is typical of supersymmetry scans also for direct detection
and neutrino detection rates. 
Low $2\gamma$ rates are obtained for binos and mixed 
neutralinos, whereas higgsinos give a larger rate. This is particularly
clear at the high mass end, where the $\Omega h^2$ requirement forces most
of the allowed models to have a large higgsino fraction, and due to the
$W$ diagrams the $2\gamma$ rate is exceptionally large for TeV higgsinos
\cite{lbe1}. In this scan, the squark masses were generally quite large,
which explains why also at low masses the models with the highest
$\gamma$ rates are mainly higgsino-like.

\begin{figure}
\begin{center}
 \mbox{\subfigure[\, ]{\epsfig{file=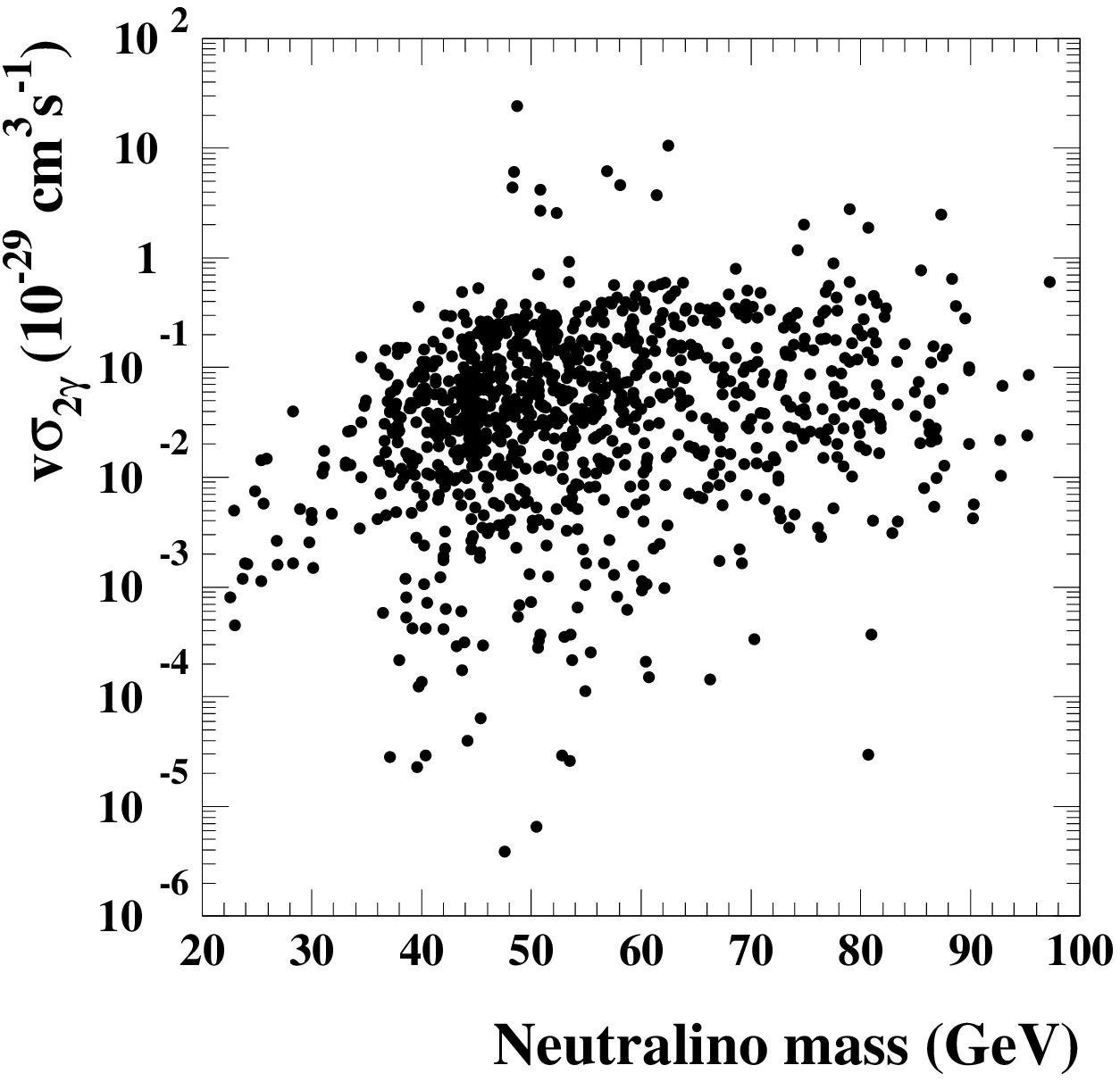,width=6cm}}\quad
 \subfigure[\, ]{\epsfig{file=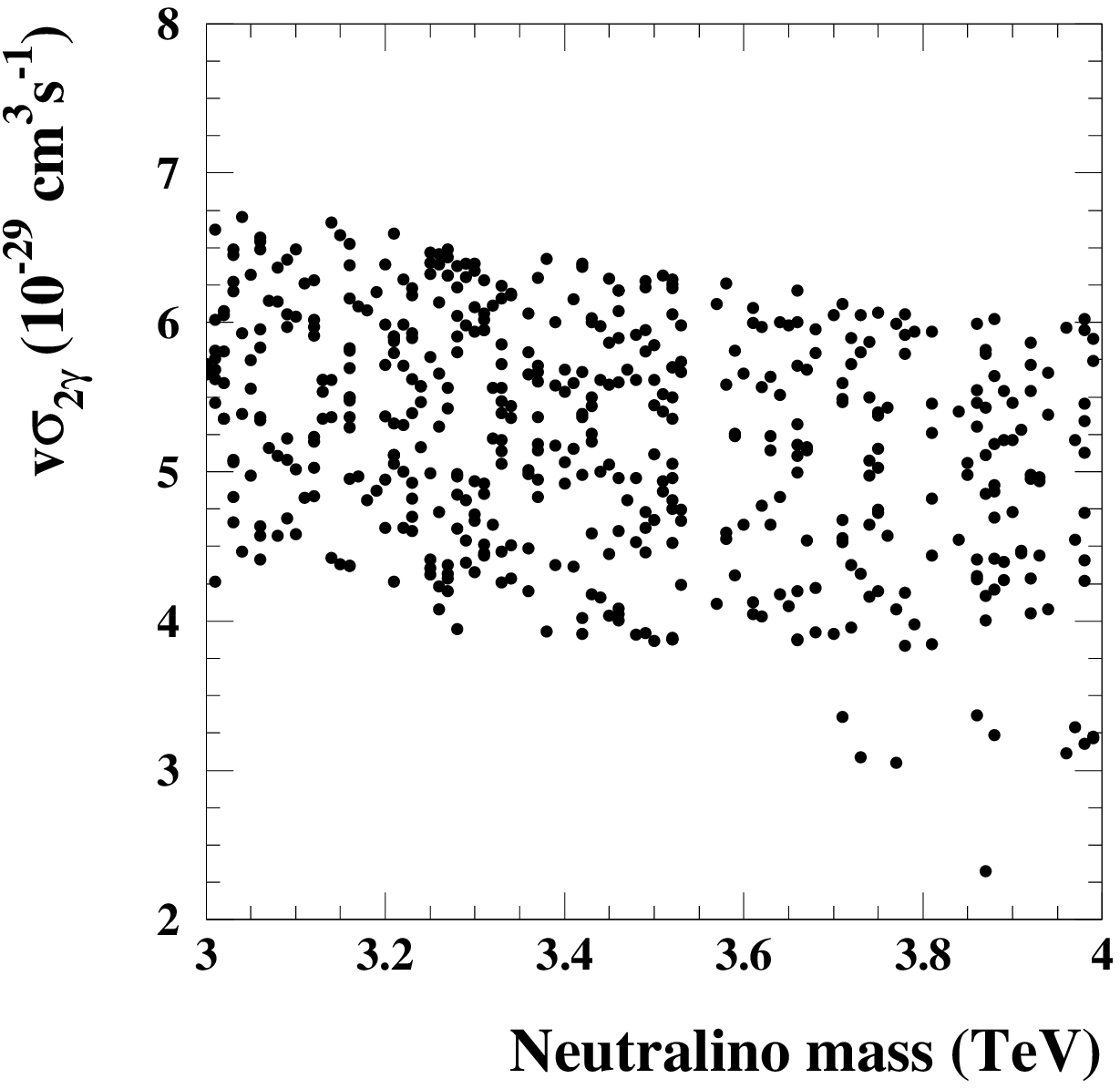,width=6cm}}}
\caption{Annihilation rate of neutralinos into 2$\;\gamma$ for more extensive 
scans in  the low- and high-mass regions.}\label{fig:fig3}
\end{center}
\end{figure}

In Figs.\,\ref{fig:fig3} (a) and (b), we have made more extensive scans
for the low- and high-mass regions respectively. In the low-mass case, 
Fig.\,\ref{fig:fig3} (a),
we have allowed a larger fraction of models with low squark mass parameter
(down to 150 GeV), since the bino rate into $2\gamma$ is dominated by
box diagrams of type (1a) and (1b). As can be seen, some models give rates
of the order of $10^{-28}$ cm$^3$s$^{-1}$. Now, because of the lower
squark masses, models which give  the highest rates
have a large gaugino component.

In the high mass region, we have focussed on the mass range between 3 and
4 TeV because of the preliminary indications in \cite{strausz} of a
possible signal. As can be seen, over that mass range there are many models 
that fulfil all accelerator and other constraints, and which all give rather
high annihilation rates. The rates are in fact much higher than the lower
limit, based on unitarity, given in \cite{lbe1} (which we reproduce).
Investigating the various contributions shown in Fig.~\ref{fig:fig2}, 
we have found that the reason for the large value of the (real part of) 
amplitude is diagram 3c, which gives a contribution proportional to 
\begin{eqnarray*}
\frac{1}{1-\frac{\mbox{\normalsize $M$}^2_{\chi^0_1}}
{\mbox{\normalsize $M$}^2_{\chi^+_i}}-\frac{\mbox{\normalsize $m$}^2_{W}}
{\mbox{\normalsize $M$}^2_{\chi^+_i}}}\;.
\end{eqnarray*}
In the pure higgsino limit, the lightest higgsino $\tilde\chi^0_1$ and 
the lightest chargino $\tilde\chi^+_1$ are almost degenerate, so when
the neutralino is much more massive than the $W$ boson this  $t$-exchange
graph becomes dominant. The approximate treatment in \cite{lbe1} only
works if $M^2_{\chi^0_1}\,/\,M^2_{\chi^+_1}$ is much smaller than unity.

\begin{figure}
\begin{center}
 \mbox{\subfigure[\, ]{\epsfig{file=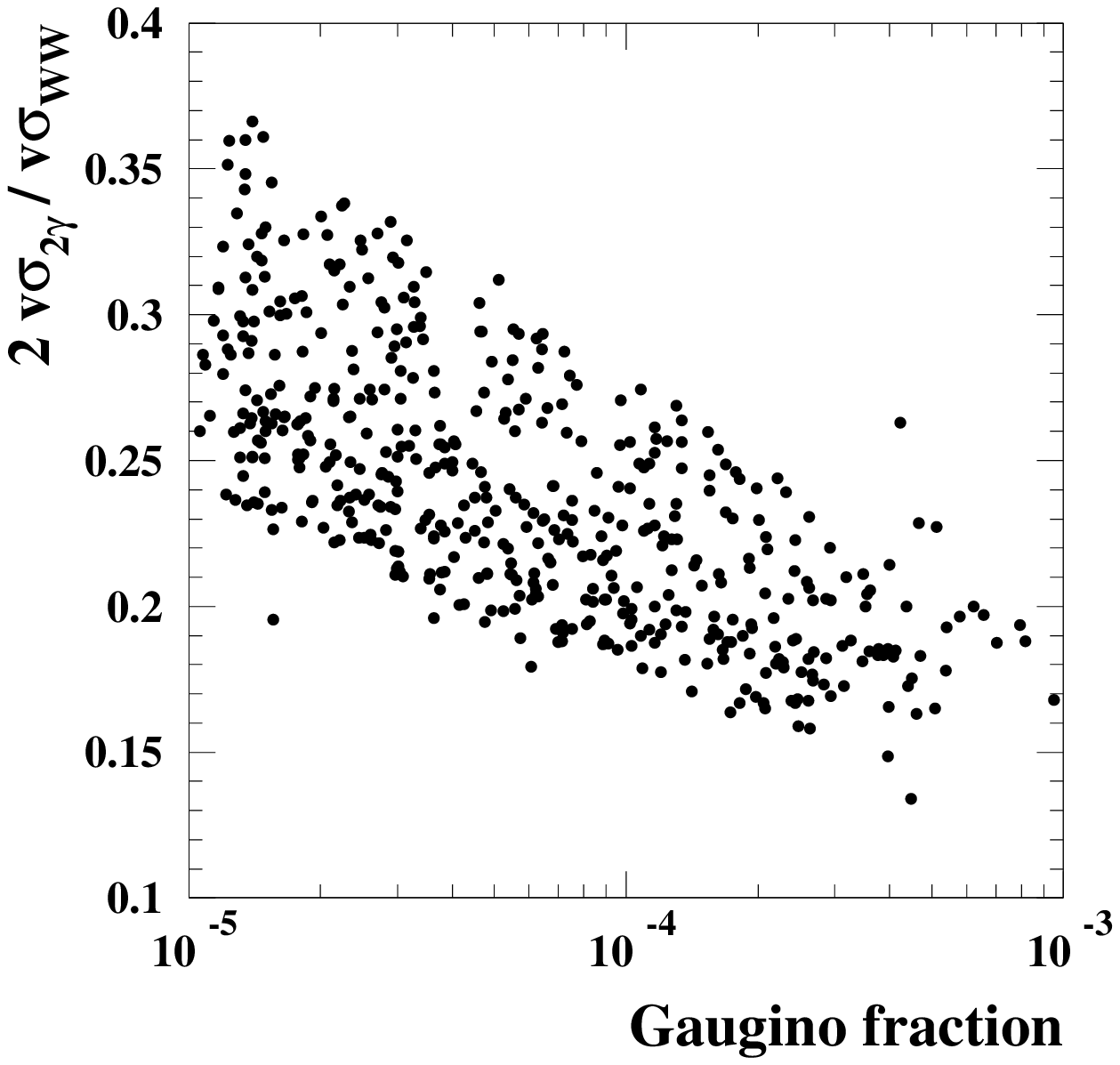,width=6cm}}\quad
 \subfigure[\, ]{\epsfig{file=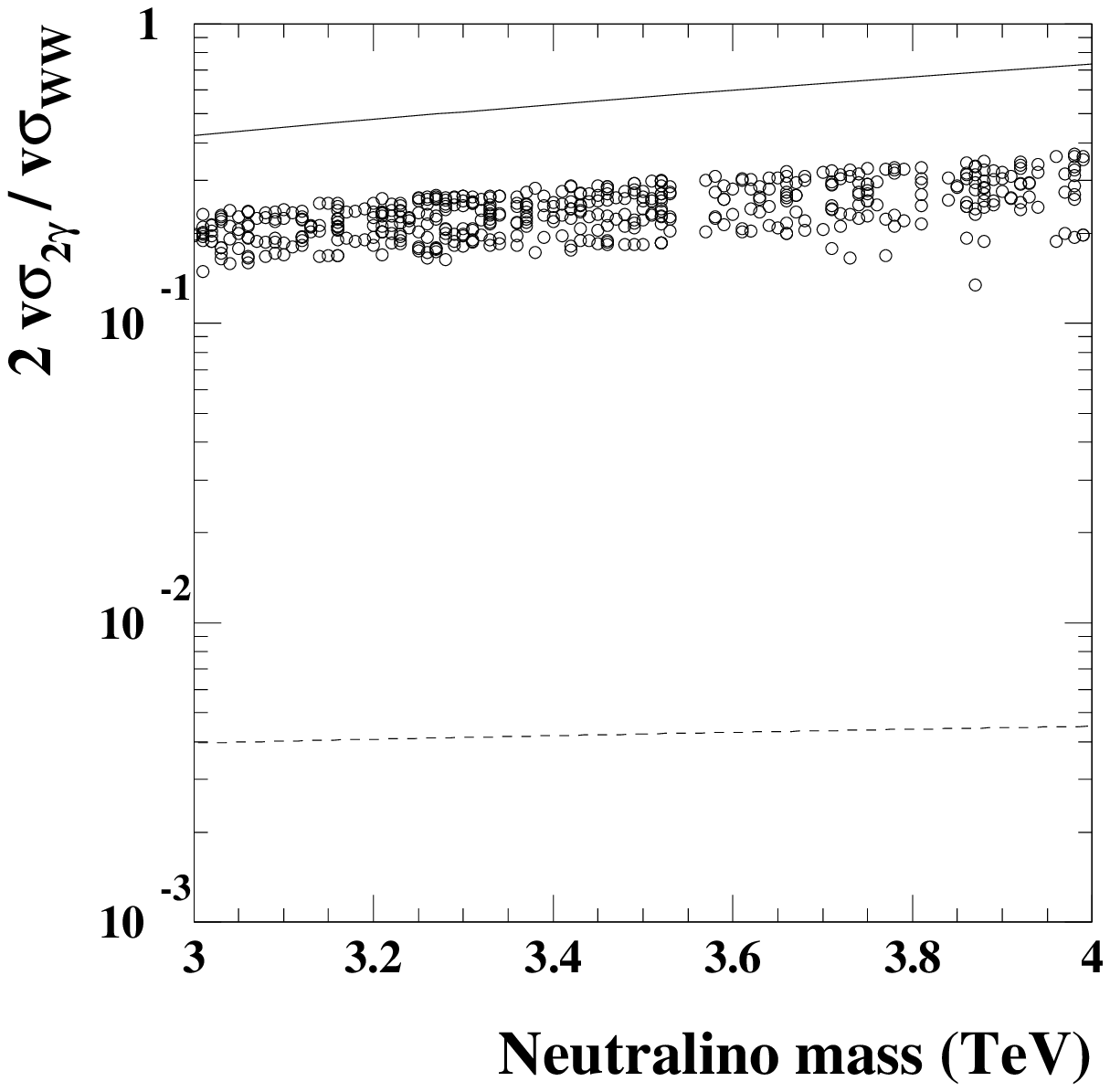,width=6cm}}}
\caption{In Fig.~(a) ratio $2\,v\sigma_{2\gamma}\,/\,v\sigma_{WW}$ 
as a function of the gaugino fraction $Z_g$; in Fig.~(b) ratio 
$2\,v\sigma_{2\gamma}\,/\,v\sigma_{WW}$ for the sampled supersymmetric 
models of Fig.~(a), the unitarity lower bound of~[20] (dashed line) and the 
pure higgsino rate obtained in this work (solid line). Adding loop 
corrections to neutralino masses [33] (not done in these plots) one may 
find points closer to the solid line.}\label{fig:fig4}
\end{center}
\end{figure}

In supersymmetric models, the mass difference between  $\tilde\chi^0_1$ and
$\tilde\chi^+_1$ goes down with increasing higgsino fraction of the 
neutralino. Thus, the $2\gamma$ rate should go up with decreasing
gaugino fraction $Z_g$. This is clearly visible in Fig.\,\ref{fig:fig4} (a).
In fact, we can approximate the pure
higgsino limit $Z_g\to 0$ by keeping only the diagrams 3a-3c and inserting 
$S_W = D_W = g_w^2/4$ ($g_w$ is the usual weak gauge coupling constant). 
Doing this, we find the roughly constant value 
\beq
v\sigma_{2\gamma}({\rm pure\ higgsino})\approx 
1.2\cdot 10^{-28}\;cm^3\,s^{-1}\label{eq:pureh}
\eeq
 over the whole range above 1 TeV.\footnote{Although we have
 not yet computed the full amplitude for $\chi\chi\to Z\gamma$ (L. Bergstr\"om 
and P. Ullio, in preparation), we have checked in the same limit that that
process also tends to a constant value of $\sigma v$ of the same order of magnitude.}

In Fig.\,\ref{fig:fig4} (b) we show the ratio 
\beq
F_\gamma\equiv{2\,\frac{v\sigma_{2\gamma}}{v\sigma_{WW}}}
\eeq
for our set of sampled supersymmetric models, bracketed by the unitarity
lower bound of \cite{lbe1} and the pure higgsino rate found here.
Notice the remarkable fact that a pure higgsino of  3 - 4 TeV mass can have
the $2\gamma$ channel as one of the dominant annihilation modes.\footnote{To be fully
consistent, this mode should then also be included in the determination
of the relic abundance. This necessitates a calculation of the 
momentum-dependence of the cross section, which has not been done. The effect
is not expected to be dramatic.}

It is an interesting coincidence that a higgsino-like neutralino in 
the 3-4 TeV range gives a value of the relic abundance which is just 
right to give closure density, $\Omega h^2\sim 0.5$. To arrive at this
conclusion, it is essential to include in the relic abundance calculation
coannihilation on the near-degenerate states, something which has only
very recently been done by Edsj\"o and Gondolo (see
Fig.\, 3 of \cite{coann}).

\section{Halo models}

In the most conservative model, the dark matter halo is described as an 
isothermal sphere with a density profile of the form:
\beq
\rho(r)=\rho_0\, \left(\frac{a^2+R^2}{a^2+r^2}\right)
\eeq
where r is the distance from the galactic centre, $\rho_0\simeq 0.3$ 
GeV/cm$^3$ 
is the local halo density, a the core radius and R $\simeq$ 8.5 Kpc is our 
galactocentric distance. The gamma ray flux in a given direction is  
\cite{turner,silk,kam}:
\begin{eqnarray}
\Phi_{\gamma}(\psi) &=&2\,\frac{v\sigma_{2\gamma}}{4\pi M_\chi^2} 
\int_{line\;of\;sight}\rho^2(l)\; d\,l(\psi) \nonumber\\
 &\simeq& 4\cdot 10^{-10} \frac{v\sigma_{2\gamma}}{10^{-29}\ {\rm cm}^3 
{\rm s}^{-1}} 
\left(\frac{10\ {\rm GeV}}{M_\chi}\right)^2\; J\left(\psi,\frac{R}{a}\right)
\,\,\, {\rm cm}^{-2}{\rm s}^{-1}{\rm sr}^{-1},
\end{eqnarray}
where $\psi$ is the angle between the direction of the galactic centre and 
that of observation. The function $J$, which is obtained performing 
analytically the integral over the line of sight, is decreasing in the 
interval [0,$\pi$] and its slope and maximum value increase as the parameter
R\,/\,a increases. A typical estimate of $J$ in the direction of the 
galactic centre is between $J(0,1/3) \sim 0.4$ and $J(0,2) \sim 2$.

Several models of the galactic halo contain the hypothesis of an enhancement
of the dark matter density in correspondence with the galactic centre, where 
there seems to be a massive nucleus. The isothermal model of Ipser and 
Sikivie~\cite{sikivie}, in which is predicted a large increase of the density 
in a region of 150 pc around the galactic centre, has been considered in 
Ref.~\cite{urban}. The estimated flux in the direction of the galactic centre 
is of the order of 10$^3\;\Phi_\gamma$, where $\Phi_\gamma\equiv
\Phi_\gamma(0)$ is the flux that would result in the conservative halo model 
choosing $J = 1$, 
and again the value is to a large extent dependent on the parameters in 
the model.

In the Berezinsky et al.\cite{berez} model, the dark matter halo density has 
the following profile:
\beq
\rho(r)=\rho_0 \left(\frac{R}{r}\right)^{1.8}
\eeq
which is assumed to be valid within 0.01 pc from the center. In this case 
the gamma ray source would appear pointlike with 99\% of the flux emitted 
concentrated in a central region defined by a solid angle of less than 
10$^{-5}$ sr. The flux in the direction of the galactic centre is highly 
enhanced, up to 10$^7\;\Phi_\gamma$\cite{urban2}.

An attractive hypothesis is that strings and texture fluctuations may have 
generated perturbations in the dark matter halo providing a clumpy density 
distribution~\cite{stebbins}. Also, in  standard CDM scenarios, smaller
``clumps'' of dark matter may merge to larger clumps like galactic halos,
and some of these objects may survive tidal forces. 
A significant increase in the gamma ray flux 
can be  obtained if typical 10$^8$ M$_\odot$ clumps are present. It is very 
difficult to detect such dark matter concentrations, nevertheless some of 
these 
may already have been identified, in the form of dwarf spheriodals residing
in the Milky Way halo. As recently as in 1994,  the Sagittarius 
dwarf~\cite{gilmore} was discovered, a dwarf spheroidal that is at a 
distance of about 
15 kpc from the centre of the Milky Way, well inside the galactic halo 
potential, and about 23 kpc from the Sun. It is dark matter dominated and 
has a density profile resembling a Heaviside function with a value of 
1.5 GeV cm$^{-3}$ on a length scale of 1 Kpc. A nice confirmation of the
hypothesis of a line signal from dark matter annihilation would be the 
detection of an increased flux in precisely the direction of 
the Sagittarius dwarf. The flux in this direction would be larger by
a factor of around 2 than the flux from the galactic centre.

Taking the uncertainty of halo models into account, we can write
the detected gamma line flux at Earth as
\beq
\Phi^0_{\gamma}=\kappa\Phi_\gamma,
\eeq
where we have seen that $\kappa$ can vary over a large range, from around 1
to at least $10^3$.

\begin{figure}
\begin{center}
 \mbox{\subfigure{\epsfig{file=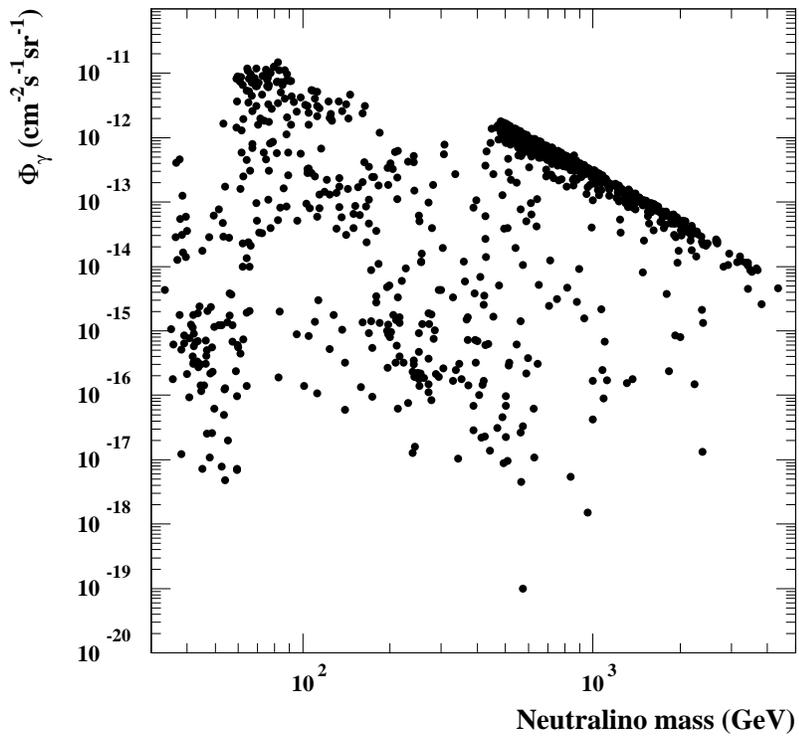,width=12cm}}}
\caption{Normalized flux $\Phi_\gamma$ for the same models as those 
in Fig.~2.}\label{fig:fig5}
\end{center}
\end{figure}

In Fig.\,\ref{fig:fig5} we show the normalized flux $\Phi_\gamma$ for
the same models as those in Fig.\,\ref{fig:fig2}.

It is clear that the flux reported in \cite{strausz}, $\sim 10^{-9}$
cm$^{-2}$s$^{-1}$sr$^{-1}$, cannot be explained by any of the scanned 
supersymmetric models in the conservative halo model. However, taking
the pure higgsino rate (\ref{eq:pureh}) in the model of Berezinsky
et al, such a high flux can be obtained towards the centre of the galaxy.
However, the (weaker) indications of a high flux in other directions
claimed in \cite{strausz} are not explicable in models where the density
is only enhanced near the galactic center. A clumpy distribution with
a large number of clumps rather smoothly distributed in the halo would
be needed to explain the high flux of \cite{strausz}.

\section{The two gluon channel}\label{sec:gluons}

\begin{figure}
\begin{center}
 \mbox{\subfigure{\epsfig{file=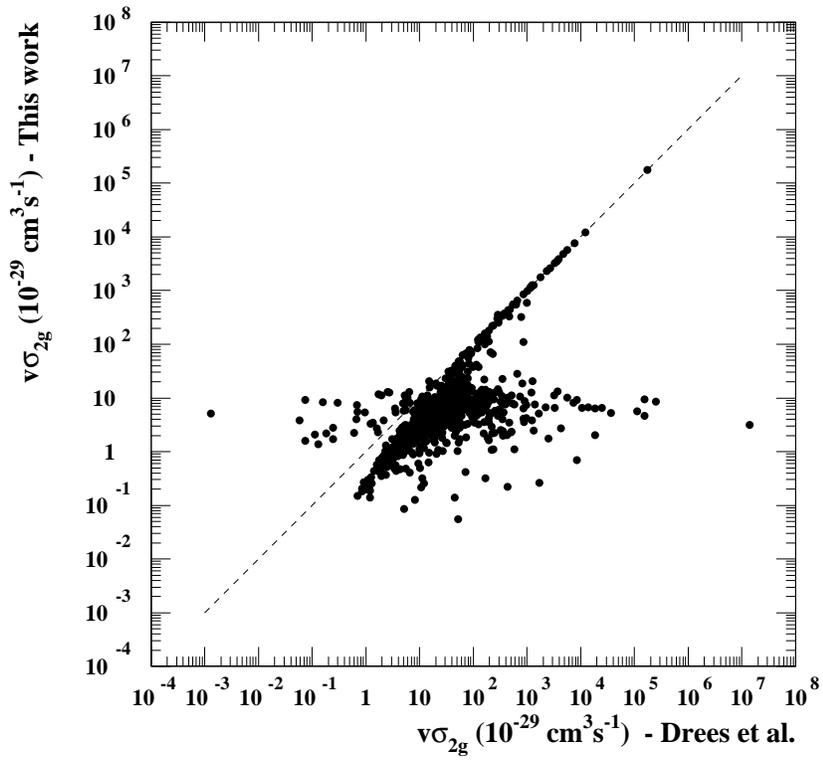,width=12cm}}}
\caption{Value of $v\sigma_{2g}$ for a scan of low-mass neutralino models 
using our results compared to the value obtained with previous 
formulas~[25].}\label{fig:fig6}
\end{center}
\end{figure}

As a final application of our loop calculation, we consider the annihilation 
channel $\chi\chi\to 2\ gluons$. Here only the diagrams 1a-1d contribute,
with of course no contribution from leptons. To get the cross section,
one just substitutes the electric charge $e_f^2$ with 1 in the expressions
(\ref{reff}) and (\ref{eq:imsquark})
 above, and the colour sum and average is performed by the
simple recipe \cite{lb1} $\alpha^2\to 2\alpha_s^2$ in (\ref{eq:sigmav}).
In Fig.\,\ref{fig:fig6} we show the value of $v\sigma_{2g}$ for the
scan of low-mass neutralino models using our results compared to those
obtained with previous formulas \cite{drees}. As can be seen, our
prediction is for most of the models below the previous ones, which could
make the indirect method of antiproton detection \cite{bottantip} 
more difficult than previously thought.

\section{Conclusions}

We have reinvestigated the process $\chi\chi\to 2\gamma$ in considerable
detail. Our new results supersede and correct previous partial results
in the literature. A new contribution is identified, which has
its main effect for heavy higgsino-like neutralinos. 
A TeV-scale pure higgsino may in fact have the $2\gamma$
final state as one of its main annihilation modes, in addition to 
giving a relic density close to the critical density of the Universe. 

A clumpy halo is needed, however, to explain the high rates indicated
in \cite{strausz} if that analysis would survive the addition of
further observational data. Irrespective of the fate of that 3.5 TeV
higgsino the gamma line signature is a promising one for neutralino detection
and should be kept in mind when designing new gamma ray detectors. For a smooth
halo, the ideal gamma ray telescope should have large collection area
($10^4$ m$^2$ or larger) and  good energy resolution
(down to percent level) but need not have good angular resolution. A large
angular acceptance would be much more important.

As a byproduct,
we have also recalculated the $\chi\chi\to 2g$ process, where we agree
with published results apart from a small, but numerically significant,
difference.

\section{Acknowledgements}
We thank J. Edsj\"o and P. Gondolo for many discussions and for 
help with the numerical calculations.
The work of L.B. was supported by the Swedish Natural Science 
Research Council (NFR).

\begin{appendix}

\renewcommand{\thesection}{Appendix \Alph{section}}
\setcounter{equation}{0}
\renewcommand{\theequation}{\Alph{section}.\arabic{equation}}

\section{ }

We define below the functions $I_{1}\left( a,b \right)$, 
$I_{2}\left( a,b \right)$ and $I_{3}\left( a,b \right)$ which appear 
in the 
formulas for the real part of the amplitude. They arise from the 
three point 
loop integrals.

\begin{eqnarray}
I_{1}\left( a,b \right) = \int_{0}^{1} \frac{d\,x}{x} \; \log \left( 
\left| 
\frac{4\,a\,x^2-4\;a\,x + b}{b} \right| \right) 
\end{eqnarray}
\begin{eqnarray}
I_{2}\left( a,b \right) = \int_{0}^{1} \frac{d\,x}{x} \; \log \left( 
\left| 
\frac{-a\,x^2+(a+b-1)\,x + 1}{a\,x^2+(-a+b-1)\,x + 1} \right| \right) 
\end{eqnarray}
\begin{eqnarray}
I_{3}\left( a,b \right) = \int_{0}^{1} \frac{d\,x}{x} \; \log \left( 
\left| 
\frac{-a\,x^2+(a+1-b)\,x + b}{a\,x^2+(-a+1-b)\,x + b} \right| \right) 
\end{eqnarray}

$I_{1}\left( a,b \right)$ is the well known function that appears in 
the 
calculation of the real part of a triangle diagram. It is possible to 
perform 
the integral over the Feynman parameter $x$, obtaining the analytical 
form:

\begin{eqnarray}
I_{1}\left( a,b \right) = \left\{
\begin{array}{ll} 
-2 \, \left( \arctan \frac{\mbox{\normalsize 
$1$}}{\sqrt{\mbox{\normalsize $b$}/\mbox{\normalsize $a$}
\mbox{\normalsize $-$}\mbox{\normalsize $1$}}} 
\right)^2
 & \;\mbox{\normalsize if}\;\; b\geq a \\[4ex]
\frac{1}{2}\, \left[ \left( \log \frac{\mbox{\normalsize $1$}
\mbox{\normalsize $+$}\sqrt{\mbox{\normalsize $1$}\mbox{\normalsize 
$-$}
\mbox{\normalsize $b$}/\mbox{\normalsize $a$}}}{\mbox{\normalsize $1$}
\mbox{\normalsize $-$}\sqrt{\mbox{\normalsize $1$}\mbox{\normalsize 
$-$}
\mbox{\normalsize $b$}/\mbox{\normalsize $a$}}} \right)^2 - \pi^2 
\right] 
& \;\mbox{\normalsize if}\;\; b\leq a 
\end{array}
\right.
\end{eqnarray} 

$I_{2}\left( a,b \right)$ and $I_{3}\left( a,b \right)$ can be 
expressed in 
terms of dilogarithms. $Li_2 (x)$ is defined according to 
Lewin~\cite{lew}. 
We find: 

\begin{eqnarray}
I_{2}\left( a,b \right) &=& -\, Li_2 \left( 
\frac{1-a-b-\sqrt{\Delta_1}}{2}
\right) \; -\, Li_2 \left( \frac{1-a-b+\sqrt{\Delta_1}}{2} \right) 
\nonumber \\
&& +\, Li_2 \left( \frac{1+a-b-\sqrt{\Delta_2}}{2} \right)\; +\, Li_2 
\left( \frac{1+a-b+\sqrt{\Delta_2}}{2}\right) \label{i2}
\end{eqnarray}

\begin{eqnarray}
I_{3}\left( a,b \right) &=& -\, Li_2 \left( 
\frac{b-a-1-\sqrt{\Delta_1}}{2\,b}
\right) \; -\, Li_2 \left( \frac{b-a-1+\sqrt{\Delta_1}}{2\,b} \right) 
\nonumber \\
&& +\, Li_2 \left( \frac{b+a-1-\sqrt{\Delta_2}}{2\,b} \right) \; +\, 
Li_2 
\left( \frac{b+a-1+\sqrt{\Delta_2}}{2\,b} \right) \label{i3} 
\end{eqnarray}

where we have defined:
\begin{eqnarray*}
\Delta_1\, = \, \left(a+b-1 \right)^2 +4\,a\; , & \Delta_2\, = \, 
\left(b-a-1 
\right)^2 -4\,a
\end{eqnarray*}

Let us consider first equation~\ref{i2}. The second member is 
explicitly real.
$\forall a , b \geq 0$, we find that $\Delta_1 \geq 0$ and that the 
arguments 
of the first two dilogarithms $(1-a-b\pm\sqrt{\Delta_1})/2$ are real 
and 
$\leq 1$: the corresponding dilogarithms are therefore real. If 
$\Delta_2 \geq
0$, $\forall a , b \geq 0$, the argument of the third and forth 
dilogarithm 
$(1+a-b\pm\sqrt{\Delta_2})/2$ are again real and $\leq 1$. If 
$\Delta_2 < 0$, 
they are complex conjugated and we can use the identity
\begin{eqnarray*}
Im \left( Li_2 ( z ) + Li_2 ( z^{\ast} ) \right) = 0
\end{eqnarray*}
to show that there is no imaginary part. In the latter case it is 
easier to 
compute numerically $I_{2}\left( a,b \right)$ introducing $Li_2 
(r,\theta)$, 
the real part of the dilogarithm of a complex argument~\cite{lew}. We 
rewrite 
equation~\ref{i2} as:
\begin{eqnarray}
I_{2}\left( a,b \right) &=& -\, Li_2 \left( 
\frac{1-a-b-\sqrt{\Delta_1}}{2}
\right) \; -\, Li_2 \left( \frac{1-a-b+\sqrt{\Delta_1}}{2} \right) 
\nonumber \\
&& +\,2\, Li_2 \left(r(a),\theta(a,b) \right)
\end{eqnarray}
where $r(a)=\sqrt{a}$ and $\cos\theta(a,b)=(1-a-b)/2\,\sqrt{a}$.

It is not necessary to implement numerically equation~\ref{i3}. 
Exploiting
Landen's functional equation~\cite{lew}:
\begin{eqnarray*}
Li_2 ( z ) + Li_2 \left(\frac{-z}{1-z}\right) = 
-\frac{1}{2}\,\log^2(1-z)
\end{eqnarray*}
it is possible to express $I_{3}\left( a,b \right)$ in terms of $I_{2}
\left( a,b \right)$. We have found the following identity:
\begin{eqnarray}
I_{3}\left( a,b \right) &=& -\,I_{2}\left( a,b \right) + 
\log^2\left(\frac{1+a+b+\sqrt{\Delta_1}}{2\,\sqrt{b}} \right) - 
\log^2\left(\frac{1-a+b+\sqrt{\Delta_2}}{2\,\sqrt{b}} \right)
\end{eqnarray}
For $\Delta_2 < 0$, the second logarithm squared has a complex 
argument and takes the form:
\begin{eqnarray*}
-\,\log^2\left(\frac{z}{|z|} \right)\,= -\, \log^2\left(\frac{|z|\,
\exp\,(i\,\varphi)}{|z|} \right)\,=\,\varphi^2
\end{eqnarray*}
where $\varphi = \arctan ((1+b-a)/\sqrt{-\Delta_2})$ .

\end{appendix}

\pagebreak

\end{document}